\documentclass[journal]{IEEEtran}
\usepackage[hyphens]{url}
\usepackage{graphicx}
\usepackage{subfig}
\usepackage{balance}
\usepackage{hyperref}

\begin{document}
\title{Audio-Based Search and Rescue with a Drone:\\ Highlights from the IEEE Signal Processing Cup 2019 Student Competition}
\author{Antoine~Deleforge,~\IEEEmembership{Member,~IEEE,}
        Diego~Di~Carlo,~\IEEEmembership{Student Member,~IEEE,}
        Martin~Strauss,~\IEEEmembership{Student Member,~IEEE,}
        Romain~Serizel,~\IEEEmembership{Member,~IEEE,}
        Lucio~Marcenaro,~\IEEEmembership{Member,~IEEE}
\thanks{A. Deleforge and R. Serizel are with Université de Lorraine, CNRS, Inria, LORIA, F-54000 Nancy, France (e-mails: antoine.deleforge@inria.fr, romain.serizel@loria.fr).}
\thanks{D. Di Carlo is with Université de Rennes, Inria, CNRS, IRISA, France (e-mail: diego.di-carlo@inria.fr).}
\thanks{M. Strauss is with Friedrich-Alexander University of Erlangen-Nürnberg, Germany (e-mail: martin.strauss@fau.de).}
\thanks{L. Marcenaro is with the Department of Electrical, Electronics and Telecommunication Engineering and Naval Architecture, University of Genoa, Genova, Italy (e-mail: lucio.marcenaro@unige.it).}
}
\markboth{Accepted to IEEE Signal Processing Magazine (Author preprint, June 2019)}%
{Shell \MakeLowercase{\textit{et al.}}: Bare Demo of IEEEtran.cls for Journals}


\maketitle



\IEEEpeerreviewmaketitle

\section{Introduction}
\begin{figure}[t!]
	\centering

	\includegraphics[width=.8\columnwidth]{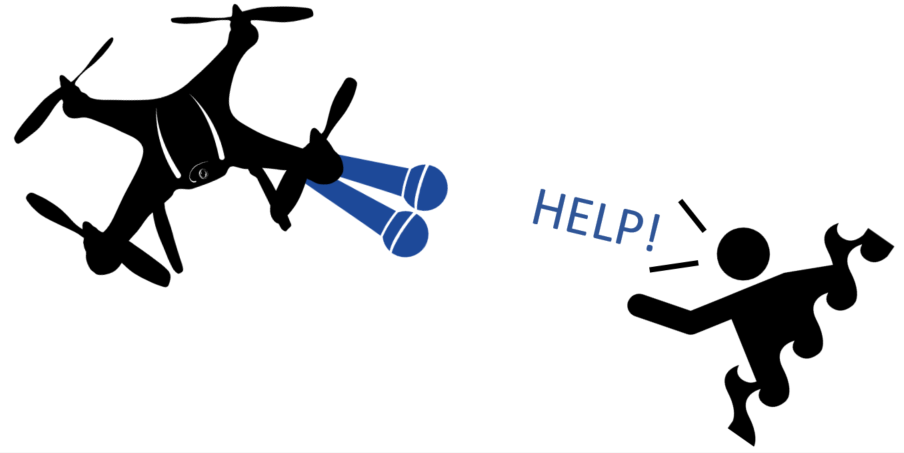}
    \caption{Microphones embedded in a UAV may help localizing people for search and rescue in disaster areas.}
    \label{fig:problem}
\end{figure}

Unmanned aerial vehicles (UAV), commonly referred to as drones, have raised increasing interest in recent years. Search and rescue scenarios where humans in emergency situations need to be quickly found in areas difficult to access constitute an important field of application for this technology. Drones have already been used by humanitarian organizations in places like Haiti and the Philippines to map areas after a natural disaster, using high-resolution embedded cameras, as documented in a recent United Nation report \cite{un2014unmanned}.  While research efforts have mostly focused on developing video-based solutions for this task \cite{lopez2017cvemergency}, UAV-embedded audio-based localization has received relatively less attention \cite{basiri2012SSL,ohata2014improvement,hoshiba2017design,wang2018tracking,strauss2018dregon}. Though, UAVs equipped with a microphone array could be of critical help to localize people in emergency situations, in particular when video sensors are limited by a lack of visual feedback due to bad lighting conditions (night, fog, etc.) or obstacles limiting the field of view (Fig.~\ref{fig:problem}).

This motivated the topic of the 6th edition of the IEEE Signal Processing Cup (SP Cup): a UAV-embedded sound source localization challenge for search and rescue. The SP Cup is a student competition in which undergraduate students form teams to work on real-life challenges. Each team should include one faculty member as an advisor, at most one graduate student as a mentor, and at least three but no more than ten undergraduate students\footnote{An undergraduate student is a student without a 4-year university degree at the time of submission.}. Formed teams participate in an open competition, and the top three teams are selected to present their work at the final stage of the competition. This year, the final took place at the 2019 IEEE International Conference on Acoustics, Speech, and Signal Processing (ICASSP) in Brighton, UK, on May 13.

In this article, we share an overview of the IEEE SP Cup experience including the competition tasks, participating teams, technical approaches and statistics.
\section{Drone-Embedded Sound Source Localization}
Estimating the direction of a sound source given audio measurements from an array of two or more microphones is a long standing research topic referred to as sound source localization (SSL) \cite{rascon2017review}. The most common approach to this problem is to estimate the sound time difference of arrival (TDOA) in a microphone pair, which can be approximately mapped to an angle of arrival when the source-to-microphones distance is large compared to the inter-microphone distance. For arrays containing more than two microphones and with known geometry, the angles of arrival of different pairs can be combined to estimate the 2D (azimuth, elevation) direction of arrival of the source in the array's coordinate frame. A large number of methods for robustly estimating TDOAs from signal pairs in the presence of noise, reverberation and/or interfering sources have been developed, including generalized cross-correlation methods \cite{knapp1976generalized} and subspace methods \cite{argentieri2007broadband,nakamura2009GEVD}. Alternatively, a number of machine-learning-based SSL methods have recently emerged, \textit{e.g.},  \cite{deleforge2015acoustic,gaultier2017vast}. However, because acquiring large enough real-world datasets to train SSL models for specific arrays is very costly, most learning-based approaches rely on simulated datasets \cite{gaultier2017vast}, which do not always generalize well to real-world conditions.

The specific task of UAV-embedded SSL comes with a number of challenges. One major issue is the noise produced by the UAV itself, generically referred to as \textit{ego-noise} in robotics \cite{lollmann2014challenges}. Due to the quickly changing speed of motors to stabilize the vehicle in the air or to change in its position, the noise profile is harmonic but also non-stationary. Additionally, since the microphones are mounted on the drone itself, they are very close to the noise sources leading to high noise levels. Because of this, the signal-to-noise ratio (SNR) can easily reach -15\,dB or less making SSL very difficult. Another factor impacting localization performance is wind noise. The wind is produced by the rotating propellers, the UAV movement in the air and may also occur naturally in outdoor scenarios. This wind noise has high power and is of low-frequency. Hence, it easily overlaps with speech signals which typically occur in a similar frequency range. Last, SSL must be performed using relatively short time windows, due to the fast movements of the UAV relative to potential sound sources. All these challenges need to be tackled at the same time and nearly in real-time when considering the real-world SSL application of search and rescue.

On the bright side however, UAV may be equipped with other embedded sensors (gyroscope, motor controllers, inertial measurement unit, compass, cameras ...) which may provide useful additional information. In the particular case of the SP Cup, the rotational speed of each of the drone's propeller at all time were provided along with the audio recordings and could be optionally used by the participants for ego-noise estimation.

\subsection{The DREGON Dataset}
\begin{figure}[t!]
	\centering
	
	\includegraphics[width=.8\columnwidth]{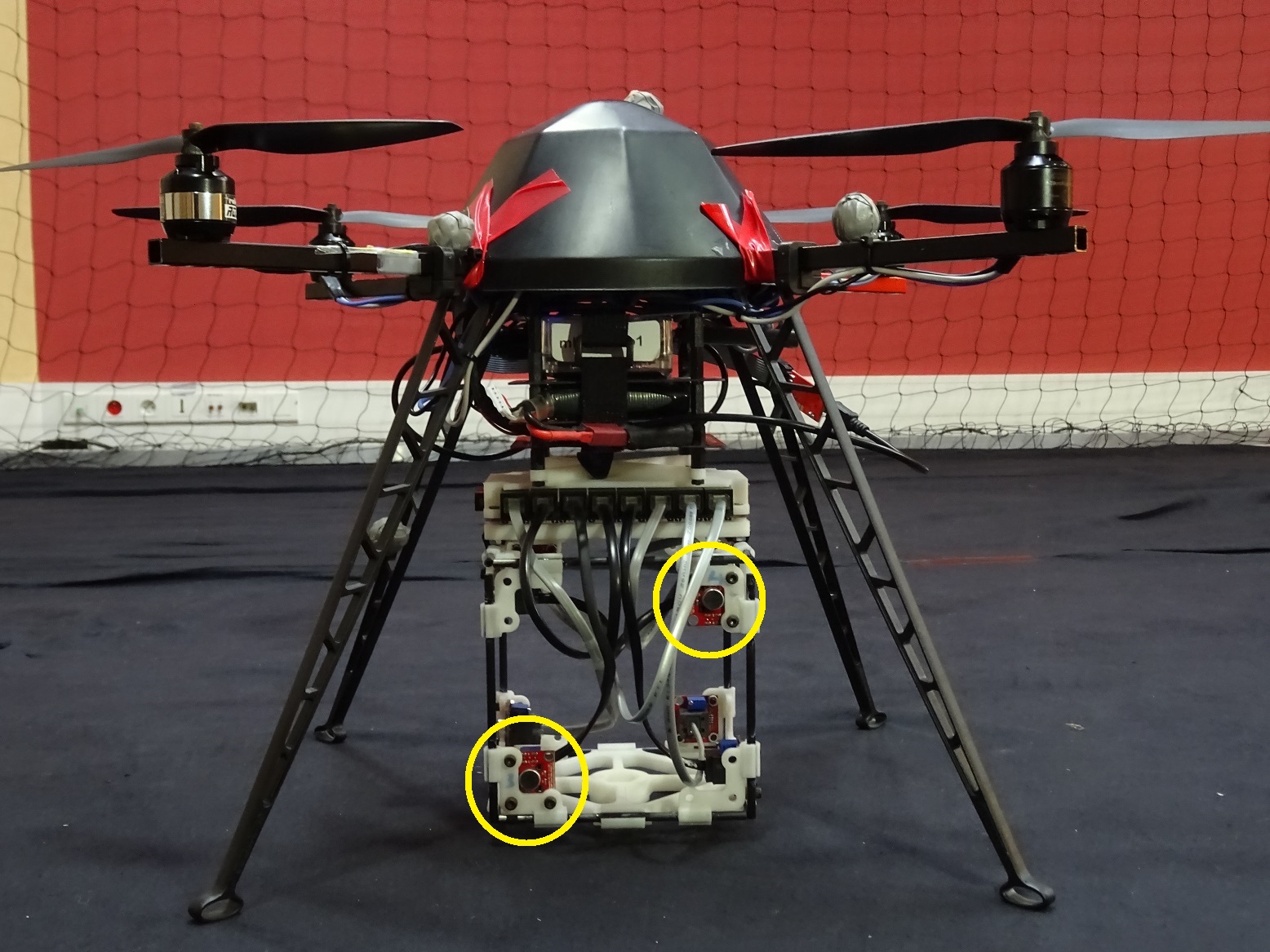}
	\caption{The quadrotor UAV used for the SP Cup, equipped with a 3D-printed 8-microphone array. Circles highlight two of the microphones.}
	\label{fig:uav}
\end{figure}

The SP Cup data were built from a subset of the recently released DRone EGO-Noise (DREGON) dataset\footnote{The release of parts of the DREGON dataset - including ground truth annotations - was delayed in order for the SP Cup to take place in fair conditions. The entire dataset is now publicly available for download at \url{https://dregon.inria.fr}.} \cite{strauss2018dregon}. It consists of annotated audio recordings acquired with a specifically designed 3D-printed cube-shaped 8-microphone array\footnote{The \textit{8SoundUSB} microphones and sound card designed by Sherbrook University (Canada) were used. Specifications can be found at \url{https://sourceforge.net/p/eightsoundsusb/wiki/Main_Page/}.} rigidly attached under a quadrotor UAV\footnote{Quadrotor UAV MK-Quadro from MikroKopter (HiSystems GmbH, Moormerland, Germany).}, as shown in Fig.~\ref{fig:uav}. It includes both static and in-flight recordings, with or without the presence of a sound source emulated by a loudspeaker emitting speech from the TIMIT dataset \cite{TIMIT} or white noise. Recordings were made inside large rooms with mild reverberation times (under 150\,ms) and negligible background noise. The synchronized 6 degrees-of-freedom coordinates of both the UAV and loudspeaker were obtained using a VICON motion capture system, yielding ground truth source direction annotation errors under $2^\circ$ for the whole dataset.

\section{Tasks in the SP Cup 2019}
The goal of the competition was for teams to build a system capable of localizing a sound source based on audio recordings made with a microphone array embedded in a UAV. Teams had to use their signal processing expertise to process the audio signals in order to extract relevant spatial cues to estimate the direction of arrival of a speech source. Key challenges are the large amount of noise present in the recordings due to the UAV's rotors and wind, and the dynamics of realistic flights, involving fast movements. To help noise estimation, the mean rotational speeds of each of the four propellers were provided for each localization task. The microphone array geometry and coordinate frame were also available.

\subsection{The Open Competition - Static Task}
For this task, 300 8-channel audio recordings at 44.1\,kHz and of roughly 2 seconds each were provided in the form of wav files. All recordings were obtained by adding together a clean recording of a static loudspeaker emitting random utterances from the TIMIT database \cite{TIMIT} from an unknown (azimuth, elevation) direction in the UAV microphone array’s frame, and a recording of UAV noise of the same length in various flight conditions and using various signal-to-noise ratios, from -20 to 5\,dB. 
The goal of this task was to retrieve the azimuth and elevation angles of the static speech source for each of the 300 recordings. A source was considered correctly localized when the great-circle distance between the estimated and ground truth directions was less than $10^\circ$. 1 point was given for each correctly localized static file, for a total of 300 points.

\subsection{The Open Competition - Flight Task}
For this task, 36 8-channel audio recordings at 44.1\,kHz lasting precisely 4 seconds each were provided in the form of wav files. All recordings were made during flight. In the first 16 recordings the source was a loudspeaker emitting speech while in the last 20 recordings the source was a loudspeaker emitting white noise. A white noise source is considered easier to localize because it has a much broader frequency range than speech. The average signal-to-noise ratio was around -15\,dB. While the source (loudspeaker) was static during flights, the microphone array was moving with the drone and hence, the (azimuth, elevation) source direction in the array’s frame was constantly changing over time. For each of the 4 seconds recordings, 15 regularly-spaced timestamps were defined.
The goal of this task was to retrieve the mean azimuth and elevation angles of the source within a 500\,ms window centered on each of these timestamps, for each of the 32 recordings. Similarly to the static task, 1 point was given for each correctly localized timestamp, for a total of 540 points.

\begin{figure*}[t!]
	\centering
	\subfloat[Static speech localization task]{\includegraphics[height=5.7cm]{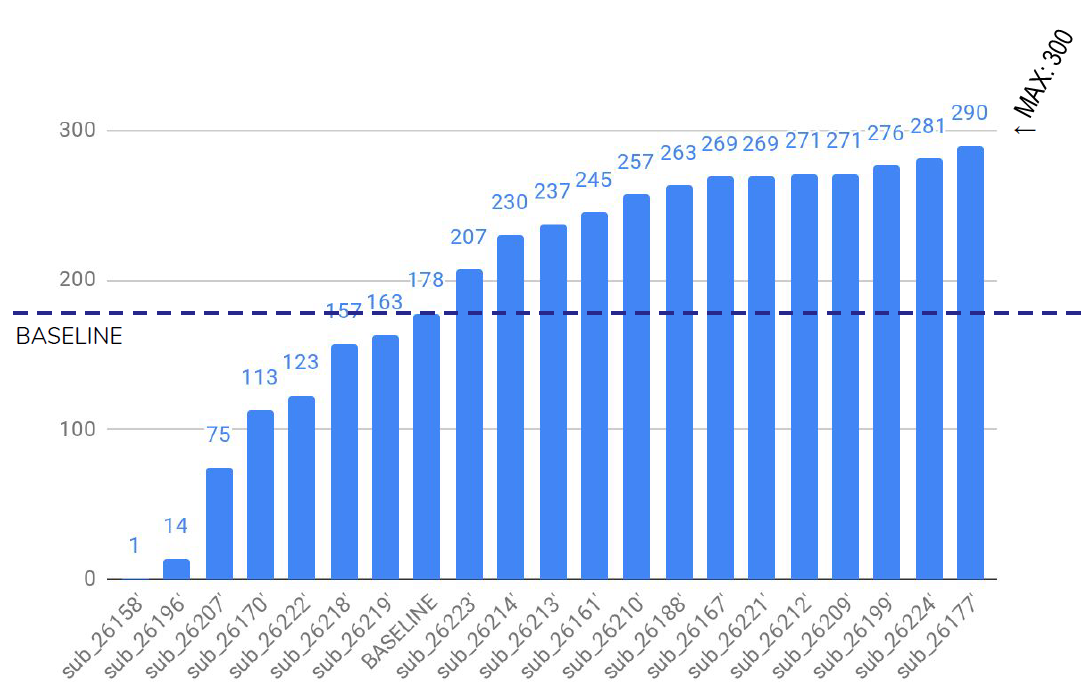}}\hfil
	\subfloat[In-flight broadband localization task]{\includegraphics[height=5.7cm]{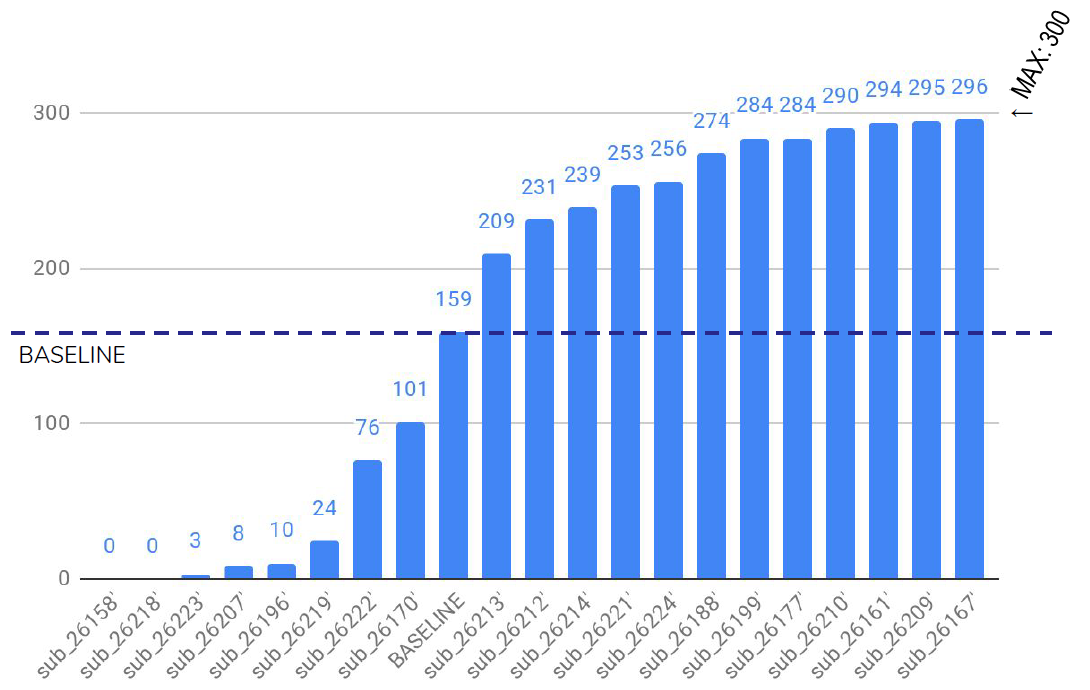}}\hfil
	\subfloat[In-flight speech localization task]{\includegraphics[height=5.7cm]{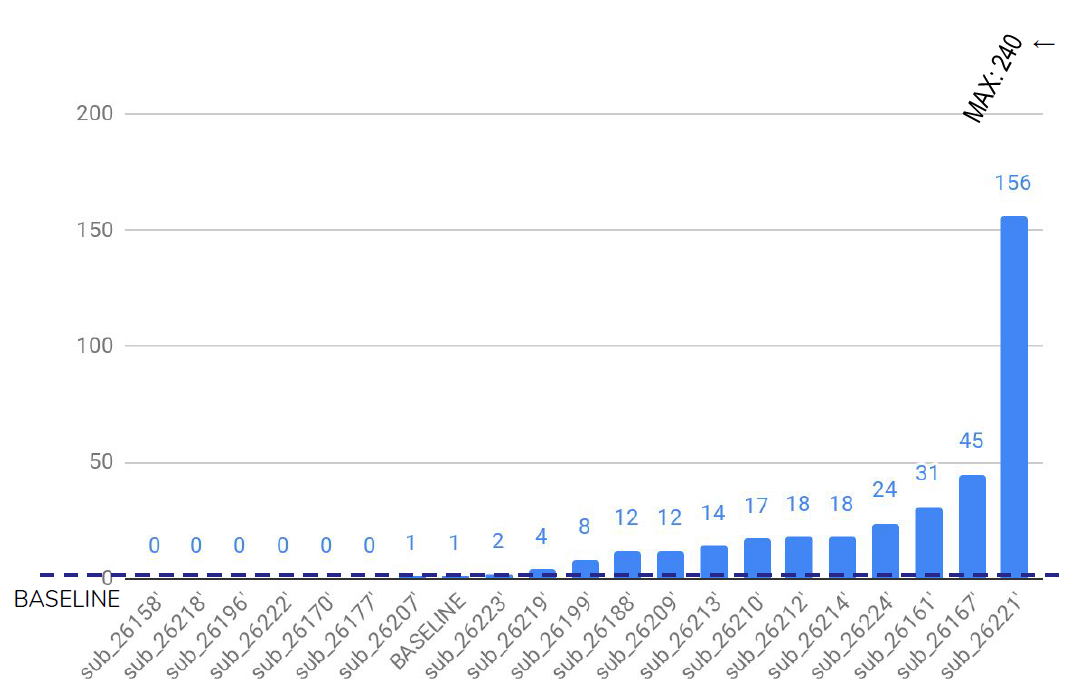}}
	\subfloat[Total scores]{\includegraphics[height=5.7cm]{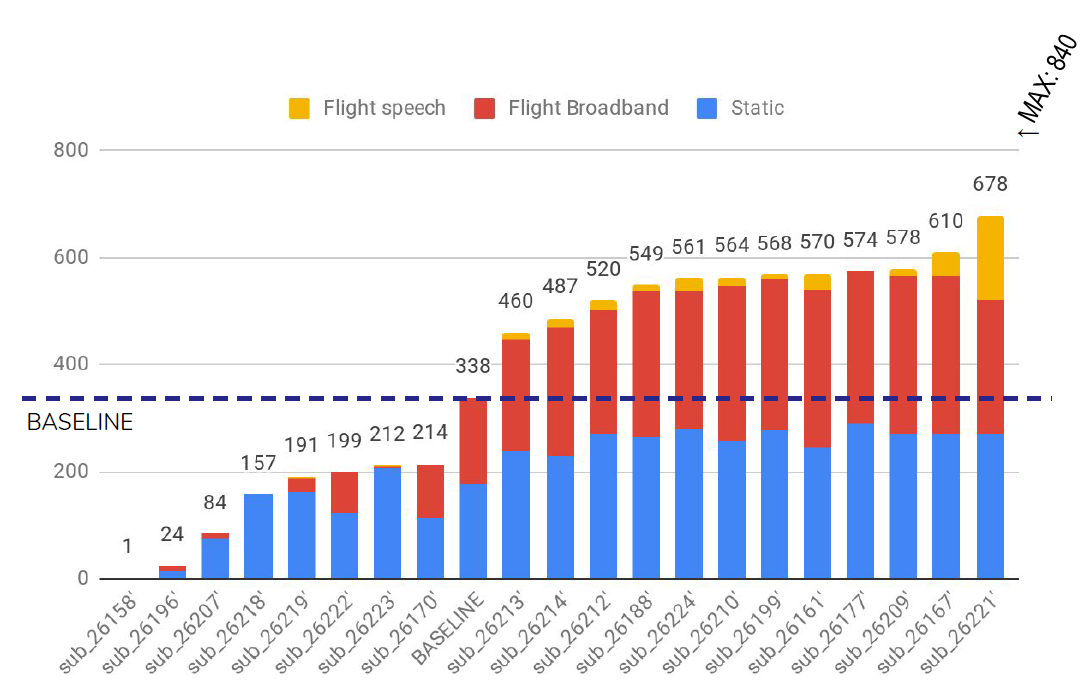}}
	\caption{Anonymised scores of the 20 teams and baseline for the three open-competition subtasks.}
	\label{fig:scores}
\end{figure*}

\subsection{The Open Competition - Bonus Task: Data Collection}
\label{subsec:bonus}
On top of the 840 points that could be gained by correctly localizing all the sources, participating teams were encouraged to send their own audio recordings obtained from one or several microphones embedded in a flying UAV. The recordings had to be made either outdoor or in an environment with mild reverberation, and should not feature other sound sources than the UAV’s noise or wind. A report detailing the microphones and UAV used, the recording conditions, and including pictures of the setup and experiments had to be given with the audio files. 60 extra points were granted to teams submitting such data.

A novel UAV ego-noise dataset was created from the teams’ contributions to this bonus task and is now freely available for research and education purpose at \url{https://dregon.inria.fr}.

\subsection{The Baseline}
A baseline method was provided for the competition. The method, implemented in MATLAB, is based on the open source MBSS Locate toolbox\footnote{\url{http://bass-db.gforge.inria.fr/bss_locate/}.} and is available on the following GitHub:
\url{https://github.com/Chutlhu/SPCUP19}. While the baseline as provided used the steered-response power method with phase transform (SRP-PHAT) method as described in \cite{lebarbenchon2018evaluation}, the MBSS Locate toolbox implements 12 different source localization methods, which are detailed in \cite{blandin2012multi}, and were also sometimes used by participants. The method chosen as baseline ranked amongst the best performing methods in the single-source localization tasks of the recent IEEE LOCATA challenge \cite{lollmann2018locata}.

\subsection{Final competition}
The three highest scoring teams from the open competition stage were selected as finalists invited to compete in the final at ICASSP 2019. Each team gave a 5 min presentation of their method followed by 5 min of questions in front of a Jury composed of SP Cup organizers and a MathWork representative. Presentations were marked by the jury according to clarity, content, originality and answers to the questions. Then, the teams were given previously unseen recordings made with the same UAV as in the open competition, namely, 20 static speech recordings of roughly 2 seconds each and one long in-flight speech recording of 20 seconds. The average SNRs for both tasks were similar to the lowest SNRs encountered during the open competition, namely, around $-17$ dB. The teams had 25 minutes to run their methods and provide results for these tasks in the same format as in the open stage. Results were evaluated on the spot, and a global score was calculated for each team so that the presentation, the static task and the flight task each accounted for one third of the total. 

\subsection{Competition Data}
The data of both the open and final stages of the SP Cup 2019 as well as corresponding ground truth result files and MATLAB scripts can be found at \url{http://dregon.inria.fr/datasets/signal-processing-cup-2019/}.

\section{SP Cup 2019 statistics and results}
As in previous years, the SP Cup was run as an online class through the Piazza platform, which allowed a continuous interaction with the teams. In total, 207 students registered for the course, and the number of contributions to the platform has been higher than 150. An archive of the class is available at: \url{https://piazza.com/ieee_sps/other/spcup2019/home}.

We received complete and valid submissions from 20 eligible teams from 18 different universities in 11 countries across the globe: India, Japan, Brasil, South Korea, New Zealand, China, Germany, Bangladesh, Australia, Poland and Belgium. The teams had 4 to 11 members for a total of 132 participants. 

Fig.~\ref{fig:scores} summarizes the scores obtained by the 20 participating teams and baseline for all the open-competition subtasks. Remarkably, 12 teams strictly outperformed the already strong baseline in overall score (excluding bonus points). As can be seen, near perfect scores were obtained by the best performing teams in the static speech and in-flight broadband tasks, with over 95\% of correctly localized sources. In contrast, the in-flight speech task which formed the heart of the competition and was the closest one to the motivational search and rescue scenario proved to be extremely challenging. For this task, only 1 timestamp out of 240 was correctly localized by the baseline. In fact, only 9 teams succeeded in localizing more than 5\% of the timestamps for this task, while the winning team achieved a stunning 65\% score.

In addition to the mandatory localization result files, 12 of the teams sent their own UAV recordings for the bonus task yielding a unique and valuable dataset (See Section \ref{subsec:bonus}).

\begin{figure*}[t!]
	\centering
	\subfloat[First place: Team AGH]{\includegraphics[height=4.2cm]{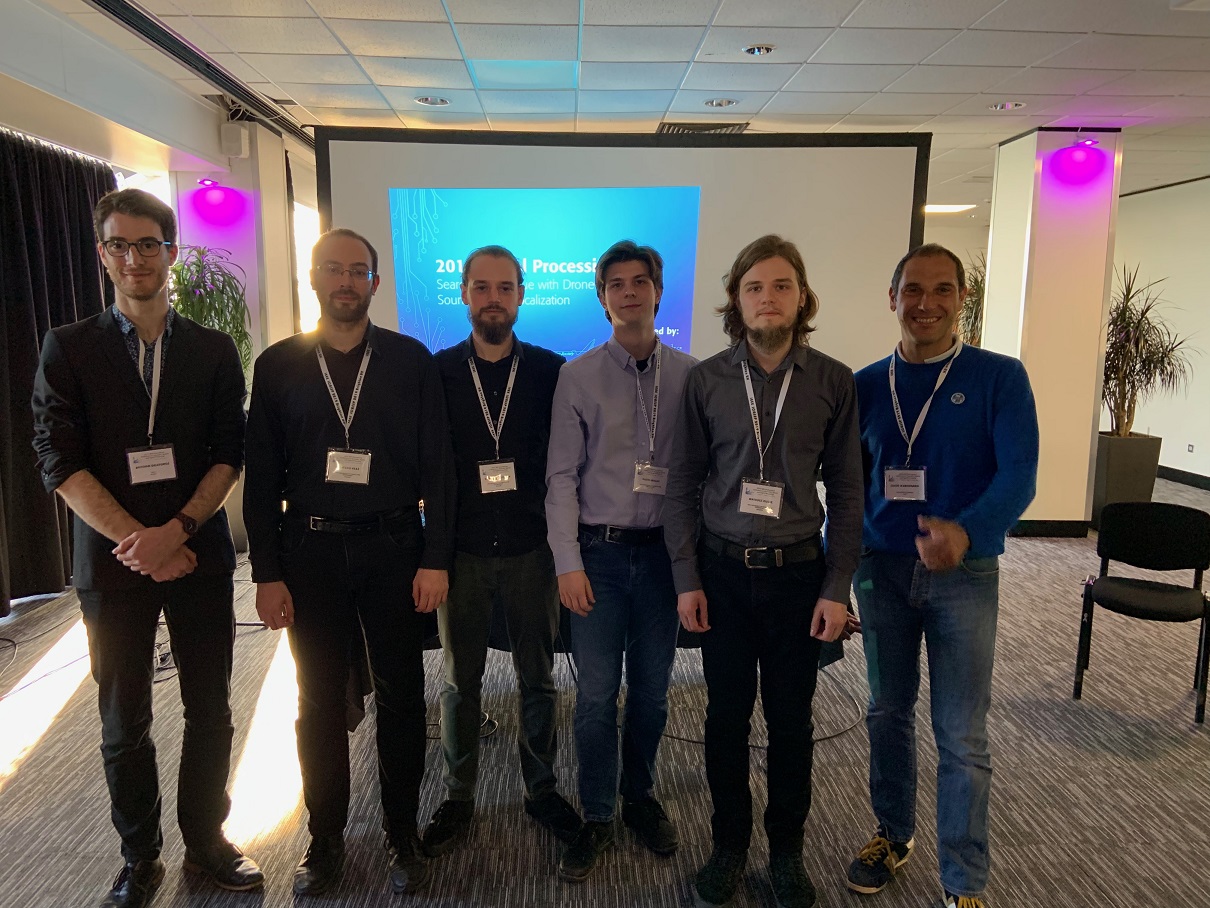}}\hfil
	\subfloat[Second place: Team SHOUT COOEE!]{\includegraphics[height=4.2cm]{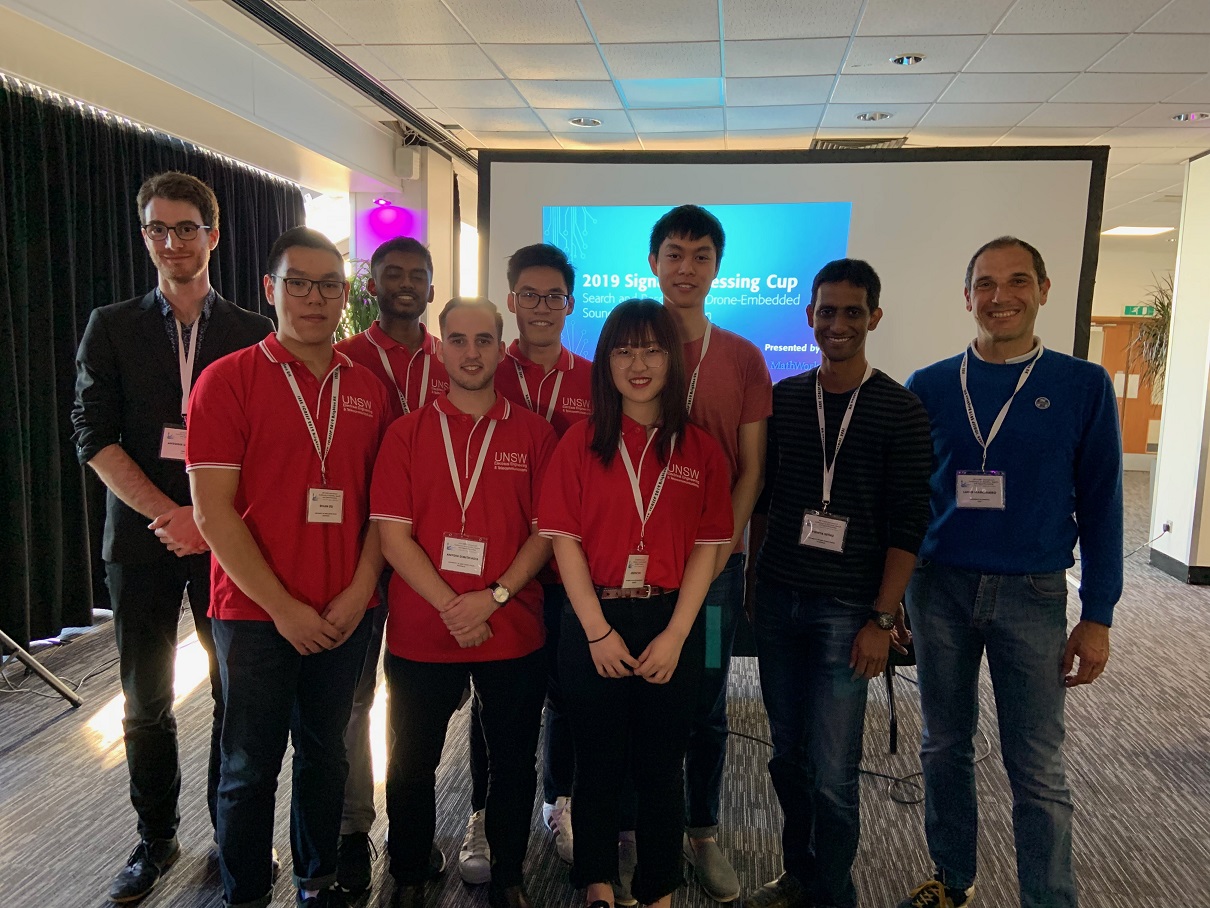}}\hfil
	\subfloat[Third place: Team Idea!\_ssu]{\includegraphics[height=4.2cm]{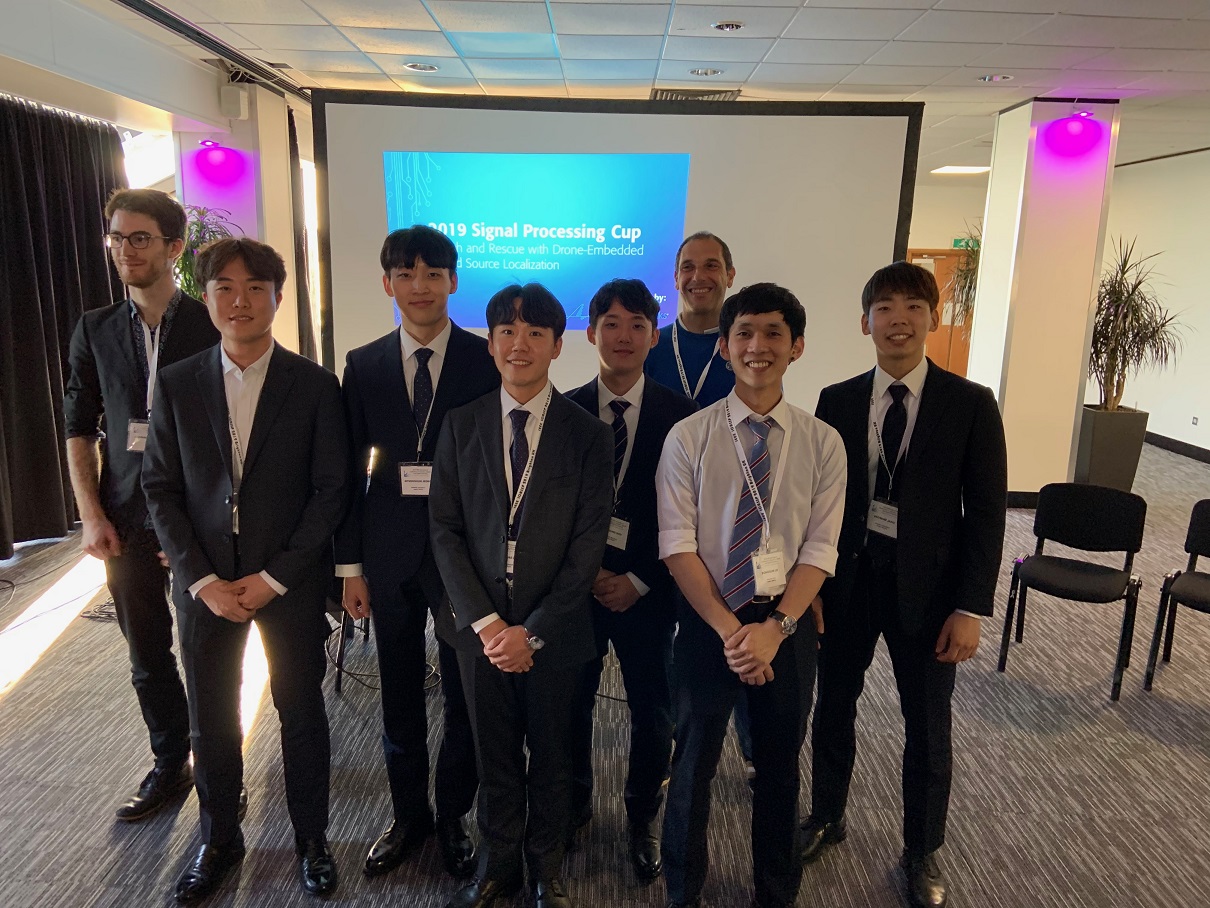}}
	\caption{Members of the three finalist teams after the final at ICASSP 2019.}
	\label{fig:teams}
\end{figure*}

\section{Highlights of the technical approaches}
In this section, we provide an overview of the methods employed by the top 12 teams which outperformed the baseline. Proposed methods were generally made of at least two components: a pre-processing stage aiming at reducing the noise in the observed signals, and a sound source localization stage.

The most popular noise reduction methods used were the multichannel Wiener filter or single channel variations of it (see \cite{gannot2017consolidated} for a review). These approaches require estimates of the noise statistics, which were obtained using many different techniques. Half of the teams used the motor speeds provided along with the audio files to do so, via some form of interpolation between the corresponding individual motor recordings available as development data. Others used voice activity detection to isolate noise-only parts and estimate the statistics on them, or made use of recursive averaging. Additionally to Wiener filtering, several teams used various bandpass filters to reduce the impact of wind noise. Notable alternatives to Wiener filtering included noise reduction methods based on nonnegative matrix factorization or deep neural networks. One team also used spatial filtering to reduce noise in the directions of the four rotors based on the provided UAV geometrical model. Two of the teams developed methods to adaptively remove microphone pairs for which the noise was too important. Many teams combined several of the above listed strategies to further reduce the noise.

For sound source localization, most of the teams built on the SRP-PHAT method implemented in the baseline. Some others used non-linear variations of it, beamforming-based methods or subspace methods. A number of teams used some form of post-processing on the angular spectra provided by these methods, for instance by ignoring regions associated to the drone's rotor directions or by clustering local minima. An approach which proved particularly successful for in-flight tasks was to smooth estimated source trajectories. This was done by using Kalman filtering or hand-crafted heuristics. Overall, the finalist teams proved that combining several techniques carefully designed for the task at hand was the only way to achieve good performance on the competition data. This suggests that even better results could be obtained by combining the best ideas from the different competitors.
\section{The Winning Teams}
In the section, we provide details about the three winning teams as well as an overview of some feedback and perspectives received from them. Pictures of the team members at the final are also reported in Figure~\ref{fig:teams}.

\subsection{Team AGH}
\subsubsection{Affiliation}
AGH (Akademia Górniczo-Hutnicza) University of Science and Technology, Kraków (Poland).
\subsubsection{Undergraduate Students}
Piotr Walas, 
Mateusz Guzik,
Mieszko Fraś.
\subsubsection{Tutor}
Szymon Woźniak
\subsubsection{Supervisor}
Jakub Gałka
\subsubsection{Approach}
The team pre-processed the signals using multichannel Wiener filtering, where the noise covariance matrices were estimated by averaging across several frames, as well as across the whole signals. To perform localization, the team combined estimates from the SRP-PHAT baseline and the GEVD-MUSIC \cite{nakamura2009GEVD} methods via K-mean clustering in the angular-spectrum domain. Angular spectra were pre-smoothed using a max filter. Finally, a Kalman filter was employed to smooth out estimated trajectories in flight tasks.
\subsubsection{Opinions}
\begin{itemize}
\item ``Leading a group of undergrads was a challenging as well as rewarding task. It gave me a perspective on how hard it is to efficiently organize research work in team, even though the team was small in number. During the competition I
especially enjoyed discussing out of the box ideas of undergrads and studying state of the art alongside them. The
tricky part of this competition was to figure out how to evaluate accuracy of tested methods, since without ground truth
you never know. On the other hand, the most exciting moments were the announcement of the results of the first phase and incontestably taking part in the final at ICASSP. This kind of competition gives an excellent opportunity for undergraduates to try their hands at solving challenging research problem.'' - \emph{Szymon Woźniak}
\item ``I chose to join the Signal Processing Cup competition because I searched for a project outside of regular studies that would allow me to develop myself in the field of signal processing. During the work I got to develop state-of-the-art sound source localization methods and also had a chance to experience working in a great team. I enjoyed the most the moments when we got some improvements after testing a new idea. Unfortunately due to the lack of development data we often had to rely on our intuition in deciding between two solutions, which was the hardest part of the competition. I think those types of events are a great chance for students to get an idea of how the scientific community works and meet like-minded people from around the world.'' - \emph{Mateusz Guzik}
\item ``I chose to participate in SPCup as I saw the opportunity to create a solution which could be potentially used for helping others. During the competition the most enjoyable and exciting part was studying state of the art algorithm, merging them into one solution and observing the results. The difficulty of the competition itself was connected to the lack of development data, which made challenging to choose between different solutions. After all, I believe that the most important of taking part in this competition was the knowledge and hands-on experience which we gained.'' - \emph{Piotr Walas}
\end{itemize}

\subsection{Team SHOUT COOEE!}
\subsubsection{Affiliation}
The University of New South Wales, Kensington (Australia).
\subsubsection{Undergraduate Students}
Antoni Dimitriadis, 
Alvin Wong, 
Ethan Oo, 
Jingyao Wu,
Prasanth Parasu,
Qingbei Cheng, 
Hayden Ooi, 
Kevin Yu.
\subsubsection{Supervisor}
Vidhyasaharan Sethu.
\subsubsection{Approach}
The team pre-processed the signals using multichannel Wiener filtering, where the noise covariance matrices were estimated from linear combinations of the provided individual motor recordings, weighted according to the current propellers speed. A non-linear generalized cross-correlation method (GCC-NONLIN \cite{blandin2012multi}) was used to localize the sound source, and for flight tasks, source trajectories were smoothed using a heuristic method inspired by the Viterbi algorithm.
\subsubsection{Opinions}
\begin{itemize}
	\item ``I learnt a lot from this SPCup competition, from how directions of arrival can be determined using signal processing techniques to how a Wiener filter can be applied to reduce the noise in recordings. Furthermore, I learnt the importance of testing and validation and how it can be utilised to evaluate the effectiveness of strategies as well as determine optimal parameters to produce an algorithm that is accurate and robust. It was an intellectually stimulating and challenging experience. I really enjoyed doing research on various strategies that could be employed in producing more accurate sound source localisation results. It was always exciting whenever new strategies developed from our research led to improved performance of our system. I chose to join the competition as I have a passion for signal processing and saw this competition as an opportunity to develop my signal processing skills. Furthermore I believed I would gain a deeper understanding of how I could apply signal processing methods and techniques to solve practical, real world problems.'' - \emph{Prasanth Parasu}
	\item ``The whole experience has been unlike any other that I have been a part of and was very much worth the time spent on the competition. Much was learnt during the SPCup, including the importance of teamwork, clear communication and (particularly in our team’s case) running programs on multiple computers to ensure that we safeguard against unforeseen problems. The UNSW team were all collaborative and supportive of each other and we have grown closer as a result. The competition gave us the opportunity to challenge ourselves intellectually and gain knowledge and experience that will serve us well in the future. I'd like to thank my team members for being so awesome and particularly our team co-ordinator, who introduced us to the competition and supported us throughout the whole adventure.'' - \emph{Ethan Oo}
	\item ``It was great to work in the team SHOUT COOEE! and compete with other brilliant teams all over the world. I love the idea of solving real world problem, it’s challenging and also attractive. Thanks to the SP Cup I gained a new understanding of speech processing, it provided a good opportunity to learn about the multi-channel Wiener filter incorporated with acoustic noise statistics of the drone. We also had chances to research and play around with different DOA estimation algorithms and seek for the best. It's an exciting and unforgettable experience.'' - \emph{Jingyao Wu}
\end{itemize}

\subsection{Team Idea!\_SSU}
\subsubsection{Affiliation}
Soongsil University, Seoul (South Korea).
\subsubsection{Undergraduate Students}
Donggun Lee,
Myeonghun Jeong, 
Minjae Park,
Youngjae Lee, 
Jinyoung Son.
\subsubsection{Tutor}
Beomhee Jang
\subsubsection{Supervisor}
Sungbin Im
\subsubsection{Approach}
The team pre-processed the signals using a combination of single-channel speech enhancement techniques and multichannel Wiener filtering, for which noise statistics were estimated from noise-only segments using voice activity detection. Wind noise was also reduced by cutting frequencies below 100\,Hz. The sound source localization method used were SRP-PHAT for the static task and GCC-NONLIN for the flight tasks. To reduced outliers on flight tasks, the team used a two-step procedure: first compute a global source direction on a 4 seconds segment, second estimate directions every 250\,ms on 1 second segments by limiting the angular search space around the global estimated direction.
\section{Future Steps and Upcoming SP Cup}
The SP Cup organizing team hopes that this edition will foster research in the emerging topic of UAV-embedded audition for search and rescue, notably thanks to its unique dataset which is now publicly available. Participants of the SP Cup 2019 as well as other researchers in this field are encouraged to submit their work to the upcoming special issue of the EURASIP Journal on Audio, Speech, and Music Processing on ``Advances in Audio Signal Processing for Robots and Drones'' (Submission deadline \textbf{December 1st 2019})\footnote{For details, please visit: \url{https://asmp-eurasipjournals.springeropen.com/call-for-papers--advances-in-audio-signal-processing-for-robots-}.}. 

The seventh edition of the SP Cup will be held at ICASSP 2020.
The theme of the 2020 competition will be announced in September.
Teams who are interested in the SP Cup competition may visit this
link: \url{https://signalprocessingsociety.org/get-involved/signal-processing-cup}.

In addition to the SP Cup, the IEEE SPS also organizes the Video and Image Processing
(VIP) Cup. The third edition of the VIP cup will be held at the IEEE International Conference
on Image Processing (ICIP 2020), in Taipei, Taiwan, October 22-25.
The theme of this edition is ``Activity Recognition from Body Cameras''.
For details, visit: \url{https://signalprocessingsociety.org/get-involved/video-image-processing-cup}.

\section{Acknowledgment}
As the SP Cup 2019 organizing committee, we would like to express our warm gratitude to all of the people who made it possible, in particular the participating teams, the local organizers and the IEEE SPS Membership Board. Many thanks also go to MathWorks and its representative Kirthi Devleker who came to the final as a member of the jury. Since its inception, the SP Cup has received generous support from MathWorks, the maker of the popular MATLAB and Simulink platforms. MathWorks kindly provided funding support to the SP Cup, including travel grants and money prizes for the finalists. Finally, special thanks go to Pol Mordel, Victor Miguet, Vincent Drevelle and François Bodin from IRISA (Rennes, France), without who obtaining such valuable UAV-embedded recordings would not have been possible.
\section{Authors}

\textbf{Antoine Deleforge} (antoine.deleforge@inria.fr) is a tenured research scientist with Inria Nancy - Grand Est (France) in the team MULTISPEECH.
He serves as a member of the IEEE Audio and Acoustics Signal Processing Technical Committee (AASP TC) and of the IEEE Autonomous System Initiative (ASI). He initiated, coordinated and chaired this edition of the SP Cup, endorsed by the IEEE AASP TC and ASI.

\textbf{Diego Di Carlo} (diego.di-carlo@inria.fr) is a PhD student with Inria Rennes - Bretagne Atlantique (France) in the team PANAMA. He was in charge of the competition evaluation and wrote the baseline and evaluation scripts.

\textbf{Martin Strauss} (martin.strauss@fau.de) is currently a master student with the Friedrich-Alexander University of Erlangen-Nürnberg (Germany). He is the main author of the DREGON dataset \cite{strauss2018dregon} which served as a basis for the competition and participated in the early design of the tasks.

\textbf{Romain Serizel} (romain.serizel@loria.fr) is an associate professor at Université de Lorraine, doing research in Laboratoire lorrain de recherche en informatique et ses applications (Loria) with the team MULTISPEECH (France). He contributed to the competition design, evaluation, and presided over the final jury.

\textbf{Lucio Marcenaro} (lucio.marcenaro@unige.it) is Assistant Professor at the Department of Electrical, Electronics and Telecommunication Engineering and Naval Architecture, University of Genoa, Genova, Italy. He chairs the Student Service Committee of the IEEE Signal Processing Society supporting the SP Cup.

\balance

\ifCLASSOPTIONcaptionsoff
  \newpage
\fi

\bibliographystyle{IEEEtran}



\end{document}